\documentstyle[prl,aps]{revtex}

\begin{document}

\title
{Vortex entry conditions in type-II superconductors.
Effect of surface defects.}

\author
{ D.Yu.Vodolazov $^1 $, I.L.Maksimov $^1$, E.H.Brandt $^2$}

\address{
$^1$ - Nizhny Novgorod  University, 603600, Nizhny Novgorod \\
$^2$ - Max-Planck-Institut f\"ur Metallforschung, D-70506 Stuttgart,
Germany}

\maketitle

\begin{abstract}

The conditions for the entry of vortices into type-II superconductors
being in the Meissner and/or mixed state, are studied by both numerical
and analytical solution of the Ginzburg-Landau equations.
A modulation instability of the superconducting order parameter is
shown to occur when the kinematic momentum (or supervelocity) of the
condensate reaches a threshold value at the superconductor edge or
surface. Due to this instability, vortices start to nucleate at the
edge or surface and then penetrate deeper into the sample. It is found
that the presence of surface defects causes a noticeable drop of
the first penetration field and leads to a qualitative change of the
magnetization curve. Based on these results a simple phenomenological
model of the edge or surface barrier, taking into account the effect
of surface defects, is suggested.

\end{abstract}

\section {Introduction}  

Since the discovery of type-II superconductors until now, much
interest was devoted to the effects that surface or edge barriers
have on the magnetic and dissipative properties of these materials.
The solution of the barrier problem is closely connected with the
condition for vortex entry into type-II superconductors. The
problem of vortex penetration was widely considered in the literature
\cite{deGen,Kramer,Fink,Aslamazov,Petukhov,Kato,Kato2,Bolech,Aranson}.
The majority of these studies \cite{Kramer,Fink,Aslamazov,Aranson}
investigated the stability of the Meissner state by analyzing the
linearized Ginzburg-Landau equations (GLE). Some studies to find
the conditions for the entry of vortex semiloops into a
superconductor \cite{Petukhov} used the London model combined with the
Gibbs free energy approach. A series of works was dedicated
to direct numerical modeling of the vortex entry into type-II
superconductors \cite{Kato,Kato2,Bolech,Aranson} by numerically
solving the time-dependent Ginzburg-Landau equations .

  In the present paper a criterion of vortex entry into
type-II superconductors, being in the Meissner and/or mixed state
is formulated based on a numerical solution of
the time-dependent Ginzburg-Landau equations (TDGLE), supplemented by
an analytical study of the linearized GLE. We show that the obtained
criterion is valid for both bulk and thin-film type-II superconductors
and allows, in particular, to estimate the first-vortex entry
field $H_{s}$ for
superconductors with an ideal surface. We also investigate the
first-vortex entry field $H_{en}$ in the presence of defects
on the surface and its dependence on the size of these
defects. It is found that  $H_{en}$ may be much lower than $H_{s}$
but still exceeds the lower critical field $H_{c1}$  (if we do
not deal with granular superconductors). Besides, we investigate
the effect of the surface defects on the magnetization curve
of pin-free superconductors. On the basis of obtained
results we suggest a phenomenological theory of edge or surface
barriers for superconductors with both ideal and imperfect surface.

The paper is organized as follows. In Sec. II the vortex entry
conditions are studied in bulk superconductors on the basis of a
numerical (Sec. II A) and analytical (Sec. II B) solution of the GLE.
In Sec. III the same problem is solved for thin superconducting films.
In Sec. IV the effect of surface
defects on the vortex entry and exit, and hence on the shape of
the hysteretic magnetization curve, is considered for the example
of bulk superconductors. On the basis of the obtained results
a simple phenomenological model for an edge or surface barrier is
suggested in Sec. V.

\section {Bulk superconductors}    

Let an infinite superconducting slab of width $w$ ($0<y<w$) be
placed in a parallel magnetic field ${\bf H}=(0,0,-H)$. The
time-dependent GLE in this case are two-dimensional because
the problem is homogeneous along the coordinate $z$,
$$
\frac {\partial \Psi}{\partial t}= -\frac{1}{C} \left[(-{\rm i}\nabla
-{\bf A})^2 \Psi +\Psi (|\Psi |^ 2-1) \right] + \chi \,, \eqno (1)
$$
$$
\frac {\partial {\bf A}} {\partial t}={\rm Re} \left [\Psi^*(- {\rm i}
\nabla{\bf A}) \Psi \right]-\kappa^2 \nabla\times\nabla\times {\bf A}.
\eqno (2)
$$
Here the length is scaled in units of the coherence length
$ \xi (T) $, time in units of $\tau=4\pi\sigma_n \lambda^2 (T)/c^2 $
($\lambda$ being the London penetration length),
the vector potential $ {\bf A} $ in terms of $ \Phi_0 / (2\pi\xi) $,
where $ \Phi_0 = ch/2e $ is the quantum of magnetic flux,
$ \sigma_n $ is the normal-state conductivity,
$\kappa=\lambda/\xi$ is the  Ginzburg-Landau parameter,
$C$ is a relaxation
constant \cite{Gorkov}, $\chi$ a random ``force'' which simulates
fluctuations of the order parameter,\cite{Kato}
and Re means the real part.

The boundary conditions for Eqs.~(1,2) along the transverse
coordinate $y$ are: $ [\nabla \times {\bf A}]_z|_{y=0,w}=-H$ and
$ (-i \partial / \partial y - A_y) \Psi|_{y=0,w} =0 $.
For $ {\bf A} $ and $ \Psi $ we assume periodic boundary
conditions along the longitudinal coordinate $x$ to model a sample
of infinite length.

\subsection {Vortex entry condition - numerical calculations} 

  Equations (1,2) are solved numerically by employing a method
similar to that described in \cite{Kato}. Namely, we introduce a link
variable $U_{j}=\exp(-i \int A_{j} dx_j)$ ($x_j=x,y)$ that allows
to satisfy the gauge-invariance condition on the grid introduced by
discretizing the variables $x$ and $y$. To solve this system of
time-dependent equations we use the Euler method.

  The parameters of the film are chosen as follows: width $w=25$,
length $L=50$ (in units of $ \xi $). Our numerical analysis shows
that the results of our study are not sensitive with respect to
an increase of $L$; the reason for this is discussed below in
this Section. The Ginzburg-Landau parameter
$\kappa=2$ and the constant $C=0.5$ were chosen to minimize the
computation time. From the microscopic theory it follows that
$C=12 $ \cite{Gorkov,Ivlev}, however in \cite{Ivlev} arguments
to prefer another selection of $C$ are given.
For each change of $H$ in steps of height $\delta H=0.01$
(in units of $H_{c2}$) we compute the solution of Eqs.~(1,2).
A series of iteration steps is performed  until the magnetic moment
becomes independent of time. The random ``force'' $\chi$ is
distributed uniformly within the interval $-s \le \chi \le s$;
the obtained results did not depend on the noise amplitude $s$
provided $s \ll 1$. In our simulations we chose $s=0.001$.

During numerical solution of the Eqs.~(1,2) we studied the
condition at which the entry of vortices into the superconductor
started. It turns out that the vortex nucleation is controlled
by one single quantity: the
supervelocity ${\bf \Pi} = \nabla \phi - {\bf A} $  (in
dimensionless units)  on the sample surface. As soon as this
quantity at the sample edge, $ \Pi_x ^ {\rm edge} $,
reaches the critical value $\Pi_{cr}$, a modulation instability
of the superconducting order parameter occurs at the edge.
The quantity $\Pi_{cr}\simeq 1$ practically does not depend on
the external field for our selection of parameters.
In what follows we consider vortex entry/exit conditions near the
left $(y=0)$ edge of the sample; from the symmetry of the
problem follows that these conditions near the
opposite edge $(y=w)$ are completely analogous.

 From our numerical solution of the GLE  we calculate the
magnetization curve of a superconductor (see Fig.~1). It is seen
that the magnetization curve has a nonmonotonic serrated shape
resulting from the threshold character of the vortex entry.
Indeed, let at some value of the external magnetic field the
inequality $\Pi_x^{\rm edge} > \Pi_{cr}$ be fulfilled on the edge
of the superconductor. Then a chain of vortices with period $l$
starts to nucleate at the slab edge, see Fig.~2. As shows our
numerical calculation, the period of the vortex chain depends
on the quantity $\Delta \Pi = |\Pi_x-\Pi_{cr}|$. The greater is
$ | \Delta \Pi |$, the less is $l$,  i.e.\ the more vortices per
unit length enter a superconductor at one step. The entering
vortices reduce the value of $\Pi_x^{\rm edge} $, since the part
of $\Pi_x$ generated by an external magnetic field, $\Pi_x^{H}$,
and the part $\Pi_x^{v}$ generated by the penetrated vortices,
have opposite sign at the slab edge. The more vortices enter the
slab at a single step of the magnetic field, the less becomes
$\Pi_x^{\rm edge}$. It is therefore necessary to further raise the
magnetic field to fulfill the condition $\Pi_x^{\rm edge}>\Pi_{cr}$
again. Particularly large serrations of the magnetization can be
obtained for a mesoscopic superconductor (of size comparable with
the vortex core size $ \xi$), as it was discussed
in \cite{Bolech,Deo}. As displays our numerical analysis, an
increase of the specimen size $L$ only leads to small corrections to
the obtained  results; this is due to the fact that vortices enter
the superconductor in form of a chain, the space period of which is
much less than $L$, see Fig.~2. Besides, it turns out that the
higher is $\Delta \Pi$ the less is the time of vortex entry.

 Table 1 gives the results of our analysis of the vortex entry into
a zero-field-cooled superconductor. The quantity $\tau$ is of the
order of $10^{-12}-10^{-13} s$ and strongly depends on the
parameters of the superconductor.

Figures 3 and 4 show how $\Pi_x^{\rm edge}$ and $j_x^{\rm edge}$
depend on the external field $H$. It is easily seen that before
reaching the value $ 0.41 $ (the entry field of the first vortices),
the current density at the edge reaches the Ginzburg-Landau depairing
current density $j_{GL}$, however vortices do not enter the sample
yet. With further increase of the applied field, the current density
reaches a maximum value $j>j_{GL}$ and then slowly decreases.
As compared to this $j_x^{\rm edge}(H)$, the dependence
$\Pi_x^{\rm edge}(H)$ has different shape. Namely, up to the barrier
suppression field $H_s$ it increases almost linearly with $H$, but
then it nearly saturates to some critical value $\Pi_{cr}$.
Moreover, by comparing Figs.~1 and 3, one sees that the sharp change
in $\Pi_x^{\rm edge}(H)$ results in a sharp change of the
magnetization $M$. Besides, we see that the quantity $ \Pi_{cr}$
practically does not depend on the magnetic field $H$. Thus,
Figs.~3 to 4 demonstrate the threshold character of vortex entry,
and also the fact that the true criterion of vortex entry is the
value of the supervelocity $\Pi_x^{\rm edge}$ at the edge, and not
the current density  $j_x^{\rm edge}$ at the edge.

\subsection {Vortex entry criterion - analytical study}  

To find the quantity $\Pi_{cr}$ analytically, we consider the
stationary Ginzburg-Landau equations
$$
\Delta F/\kappa^2+F (1-F^2- {\bf \Pi} ^2) =0 \,,
$$
$$
\nabla \times \nabla \times {\bf \Pi} +F^2 {\bf \Pi} =0 \,,
$$
in which we introduced new variables $F=|\Psi|$ and ${\bf \Pi}$.
In this case it is more convenient to measure distances in units of
$\lambda$. The supervelocity $|{\bf \Pi}| $ is
in units of $ \Phi_0/2\pi\xi $, and the magnetic field in units of
$ \Phi_0 / 2\pi\xi\lambda = \sqrt{2} H_{c}$,
where  $H_{c}=\Phi_0 /\sqrt 8 \pi \lambda \xi$ is the thermodynamic
critical field \cite{deGenbook}. We seek the solution of
these equations in the form $F=F_0 +\tilde f$ and
$ {\bf \Pi}={\bf Q}+\tilde {\bf q} $, where $ \tilde f $ and
$ \tilde {\bf q} $ are small ($ |\tilde {\bf q}|  \ll | {\bf Q}| $,
$  \tilde f \ll F_0 $) perturbations of the stationary values
of the order parameter $F_0$ and momentum ${\bf Q}$.

Disregarding terms of order higher than linear in
$ \tilde {\bf q} $ and $ \tilde f $ we obtain a system of three
linear differential equations for the functions $ \tilde f $ and
$ \tilde {\bf q} $,
$$
 \frac {1} {\kappa^2} \Delta \tilde f + (1-3 F_0^2- {\bf Q} ^2)
 \tilde f -2 F_0 ({\bf Q \tilde q}) =0 \,,
$$
$$
 -\frac {\partial^2 \tilde q_y} {\partial x^2} + \frac {\partial^2
 \tilde q_x} {\partial x \partial y} +F_0^2 \tilde q_y
 + 2\tilde f F_0 Q_y=0 \,,
$$
$$
 -\frac {\partial^2 \tilde q_x} {\partial y^2} + \frac {\partial^2
 \tilde q_y} {\partial x \partial y} +F_0^2 \tilde q_x
 + 2\tilde f F_0 Q_x=0 \,.
$$
Performing the Fourier transform with respect to the longitudinal
coordinate $x$ with wavevector $k$ (physically this means we search for
the solution as a chain of vortices with a period $\sim 1/k$) we obtain
the following system of equations:
$$
 \frac{1}{\kappa^2} \frac{d^2 f}{dy^2}+f(1-3F_0^2-{\bf Q}^2
 -{k/\kappa}^2) -2F_0({\bf Qq})=0 , \eqno (3)
$$
$$
 \left (k^2 q_y+ik\frac{dq_x}{dy}\right)+q_yF_0^2+2fF_0 Q_y=0 \,,
 \eqno (4)
$$
$$
-\frac{d^2 q_x}{dy^2}+ik\frac{d q_y}{dy}+q_xF_0^2+2fF_0 Q_x=0 ,\eqno(5)
$$
supplemented by the boundary conditions: $df/dy|_{y=0,w} = 0 $,
$q_y|_{y=0,w} = 0$, and $(ikq_y-dq_x/dy)|_{y=0,w} = 0$. Hereafter,
$f$ and ${\bf q}$ denote the Fourier-transforms of the
perturbations $ \tilde f $ and $ \tilde {\bf q} $.

  As shows our numerical analysis of the equations (1,2)
(with $\kappa=2$) the perturbation is localized at the slab surface
over a length of about $ \xi $. This feature will be analytically
confirmed below for the case $\kappa \gg 1$, at which  perturbations
are also localized near the surface over a length of
several $ \xi $ (i.e. much less than $ \lambda $).
For such a distance from the surfaces one has $Q_y \approx 0 $
as a result of our boundary condition $Q_y^{\rm edge} = 0 $ and the
absence of vortices near to the surface. Besides, it is possible to
expand $Q_x $ and $F_0 $ near the surfaces of the slab into a
Taylor series (with respect to $y$) keeping only the linear terms.
It is possible to take advantage of the relation $F_0^2 = 1-{\bf Q}^2$
that follows from the GLE for the order parameter $F_0$ ( neglecting
the term with second derivative). Inserting $q_y$ from Eq.~(4)
into Eq.~(5) we obtain a system of two second-order
differential equations, valid near the slab surface:
$$
 \frac{1}{\kappa^2} \frac{d^2 f}{\partial y^2}+f(1-3F_0^2-Q_x^2
 - (k/\kappa)^2)  - 2F_0Q_xq_x=0 \,, \eqno (6)
$$
$$
 - \frac{d^2q_x}{dy^2}\frac{F_0^2}{k^2+F_0^2}+\frac{dq_x}{dy}
 \frac{d}{dy} \frac {k^2}{k^2+F_0^2} +q_xF_0^2+2fF_0 Q_x=0 \,.
      \eqno (7)
$$
In the Meissner state Eqs.~(6), (7) are valid not only on the slab
surface, but also in the bulk, since in the Meissner state one has
$Q_y=0$ everywhere in the slab. Equations (6) and (7)
have been derived and studied numerically in
Refs. \cite {Kramer,Fink} to check the stability of the Meissner
state in bulk superconductors.

We are interested in the solution of Eqs.~(6) and (7) that vanishes
deep inside the slab, choosing the surface at $y=0$. In what follows
we shall concentrate on the physically most interesting limit
$\kappa \gg 1$. This allows us to neglect the term with the first
derivative in Eq.~(7) since the function $k^2/(k^2+F_0^2) $
depends on $y$ rather smoothly (in the limit
$k \approx \kappa \gg 1$ and $F_0 \leq 1$) .
In addition, the choice $\kappa \gg 1$ allows us to disregard the
term with the second derivative in Eq.~(6) at the critical
wavenumber $k=k_c $ (see below). As a result we arrive at the
Airy equation for the perturbation $q_x$,
$$
\frac{d^2q_x}{dy^2}+q_x(b+ay)=0 \,. \eqno (8)
$$
It is worth noting here that Eq.~(8) with proper boundary conditions
is applicable to study the stability  of both Meissner state and mixed state
in  bulk superconductors with respect to the  vortex  entry .
In the latter case the validity region  of Eq.~(8) is $y_0 \ll w_{vf}$,
where $y_0$ is the size of perturbation in the $y$-direction and $w_{vf}$
is the width of the vortex-free zone near the superconductor surface.
In Eq.~(8) the coefficients $a$ and $b$ are expressed through the
constants $P$ and $R$, which appear while expanding the order
parameter $F_0^2 \approx 1-P $ and the vector potential related
quantity  $Q_x^2 \approx P+Ry $ near the superconductor edges,
$$
b=-(1+k^2-P)(6P-2-(k/\kappa)^2)/(2P-2-(k/\kappa)^2) \,,
$$
$$
a=-4R(1+k^2-P)(3P-2-(k/\kappa)^2)/(2P-2-(k/\kappa)^2)^2 \,.
$$
The coefficient $\, R=-2\sqrt P H_{\rm edge}$ is expressed through
the magnetic field near the edge, $H_{\rm edge}$, which for bulk
superconductors is equal to the external field $H$; the absence of
linear terms in the expression for $F_0^2(y)$ follows from the
boundary condition  $\partial F_0/\partial y |_{y=0}=0$. The
solution of the Eq.~(8) is the Airy function
$$
q_x=D\Phi [-a ^ {1/3} (y+b/a)] \,,
$$
where $D$ is an integration constant.

From the boundary condition $\partial q_x / \partial y|_{y=0}=0$
(derived from the conventional condition $q_y|_{y=0}=0$)
follows the simple relation
$$
b=-\gamma_1 a^{2/3}, \eqno (9)
$$
where $\gamma_1 \approx -1.02$ is the first zero of the first
derivative of the Airy function. Equation (9) defines the value of
the supervelocity $\Pi$ on the slab surface for which a periodic,
spatially modulated disturbance $q_x \not=0$ exists for a given
value of the perturbation wave vector $k$. Inserting the
expressions for $a$ and $b$ into Eq.~(9) we obtain the dispersion
relation in the form
$$
(P-1-k^2) \left (6P-2-(k/\kappa)^2 \right)^3 \left (2P-2-(k/\kappa)^2
\right) = gP \left (3P-2-(k/\kappa)^2 \right)^2 \, \eqno (10)
$$
that contains only one field-dependent control parameter
$$
g = 64 | \gamma_1 | ^ 3 H _ {\rm edge} ^2 .
$$
The given equation relates the value $ \Pi $ on the slab
surface to the wavenumber $k=k_c$ for which a nontrivial
solution $q_x$ becomes possible.

Numerical analysis of Eq.~(10) shows that the minimum
(but non-zero) value of $P(k)$ is reached for $k=k_c$. In Fig.~5
the dependence $\Pi_x^{\rm edge}(k)=\sqrt{P (k)}$ is presented
for different values of $ \kappa $ and $H $. It is important that the
value $\Pi_{cr} \equiv \sqrt{P(k=k_c)}$ for $\kappa=2$ is practically
independent of the magnetic field $H$. Indeed, for $H=0.41H_{c2}$
[$g(H=0.4H_{c2}) \approx 45$] $ \Pi_{cr}$ is equal to $ \simeq 0.78$,
while for $H \simeq H_{c2}$  [$g(H\simeq H_{c2}) \approx 280 $]
$\Pi_{cr}\simeq 0.79$ differs only slightly.
Numerical results confirming this
independence of $\Pi_{cr}$ from $H$ are presented in Fig.~3.
A slight quantitative difference is possibly due to the fact
that Eq.~(10) was derived in the limit  $\kappa \gg 1 $.
Note also that for the latter case the difference between the
critical values $\Pi_{cr}$ at the field of first vortex entry
(i.e. at $H=H_s \approx H_c$ \cite{deGen}) and at the field near
$H_{c2}$ becomes more appreciable (0.66 and 0.81, respectively, for
$ \kappa=100 $); however the difference in  $\Delta \Pi_{cr}$ does
not exceed $20\%$.

Our analysis of the stability of the Meissner state for the case
$ \kappa=2 $, performed on the basis of the linear Eqs.~(6) and (7),
gives a critical value $\Pi_{cr}=1.02 $ (for $H=0.41H_{c2}$), which is
quite close to the value $\Pi_{cr}=0.97$ produced by direct
solution of the nonlinear GL equations (1) and (2), but differs
appreciably from the result of Eq.~(10): $\Pi_{cr}=0.78$. A similar
comparison performed for the case $\kappa=5 $ gives a more
satisfactory coincidence of the $\Pi_{cr}$ values
(0.8, 0.85 and 0.75, respectively), which were
obtained by solving Eqs.~(1)-(2), (6)-(7) and (10), respectively.

It is interesting to note that for the case $\kappa \gg 1$, the
perturbation is localized at the superconductor surface; this is
reflected by the exponential decay of the Airy function
$ \Phi(\Theta)$  in the region $\Theta >0$. The estimate for the
characteristic scale $y_0$ of the perturbation decay deep into the
film yields $ y_0 \sim b^{-1/2}$. The numerical solution of (10)
shows that for large values of $\kappa$ one has
$b(k_c) \approx k_c^2 \approx 0.1 \kappa^2 $ (see Fig.~5); therefore,
we obtain finally that the perturbation is localized on a scale of
several $\xi$ along the $y$-direction. Besides, for $\kappa \gg 1$,
the important assumption $\partial^2 f/\partial y^2 \ll f \kappa^2 $
is justified for $k=k_c\approx (0.3-0.4) \kappa$, which permits to
simplify Eq.~(6).

\subsection {Vortex exit}   

Let us discuss now the conditions for the vortex exit which takes
place when the external field is decreased. As can be seen from
Figs.~1 and 3, the dependences $M(H)$ and $\Pi_x^{\rm edge}(H)$ are
nearly linear functions of the (initially) decreasing magnetic field.
Physically this is explained by the fact that during the initial stage
of the field decrease the vortices do not exit from the superconductor.
At some value of the magnetic field (in our case at $H=0.36H_{c2}$) the
vortices start to exit from the sample. At $H<0.36H_{c2}$ both dependences
$M(H)$ and $\Pi_x^{\rm edge}(H)$ become nonmonotonic because not one
but several vortices leave the sample at once. Our study shows that
the process of the vortex exit is controlled by the behavior of
$\Pi_x^{\rm edge}$: as soon as $\Pi_x^{\rm edge} $ changes sign with
the decrease of $H$, the vortices leave the superconductor. In the
opposite case the vortices do not exit.

 At first sight Fig.~3 seems not to confirm the above statement; this
is explained by the fact that only stationary values
of $\Pi_x^{\rm edge}$ are shown there. It turns out that during the
initial decrease of the magnetic field $\Pi_x^{\rm edge}$ also
decreases. For example, at $H=0.37H_{c2}$ one
has $\Pi_x^{\rm edge}=0.04$;
when the magnetic field decreases down to $H=0.36H_{c2}$,
$\Pi_x^{\rm edge}$ first becomes negative, and then, after exit of
several vortices, reaches the stationary value $0.17$.
This process repeats at lower  values of the magnetic field, e.g.\ at
$H/H_{c2}=0.25, 0.16, 0.09$ for given sample/material parameters.

  We should emphasize that when the field decreases below
some value $H=H^*$ (in our case  $H^*=0.41H_{c2}$), the magnetization
and magnetic field coincide in sign. Such a ``paramagnetic'' response,
resembling the result obtained by \cite{Tern} for the bulk case,
is essentially due to surface pinning, which is a generic feature of
type-II superconductors with an ideal surface.

\section {Thin films}  

Extending the method developed for bulk superconductors, we shall
find below the criterion for vortex entry into a thin
($d < \lambda$) superconducting film placed in a perpendicular
magnetic field. In this geometry ${\bf \Pi}$ is practically
independent of $z$; together with the boundary conditions
$\partial |\Psi |/\partial z |_{z=\pm d/2}=0$ this gives a
$z$-independent order parameter $\Psi$. By averaging equation (1)
over the coordinate $z$ one arrives at the two-dimensional GL
equation. Using the method suggested by \cite{Pearl} for thin films,
we write the GLE for $ {\bf \Pi} $ valid in all space as
  $$  
  \nabla \times \nabla \times {\bf \Pi} = -F^2 {\bf \Pi} \delta (z)
  \theta (y) \theta (w-y) \,. \eqno (11)
  $$
The equation to find the order parameter, valid for superconducting
domain $z=0$, $0\leq y \leq w$, $-\infty \leq x \leq + \infty$, reads
  $$  
  \kappa_{\rm eff}^{-2} \Delta F+F(1-F^2-{\bf \Pi}^2)=0 \,, \eqno (12)
  $$
where $\kappa_{\rm eff}=\lambda_{\rm eff}/\xi$,
$\lambda_{\rm eff}=\lambda^2/d$, and the distance now is measured
in units of $\lambda_{\rm eff}$, ${\bf \Pi} $ in units of
$ \Phi_0/2\pi\xi $, and the magnetic field in units of
$\Phi_0/2\pi \xi \lambda_{\rm eff}$. Similarly to the case of bulk
superconductors, we linearize Eqs.~(11) and (12) with respect to
small ($ |\tilde {\bf q}| \ll | {\bf Q}|$, $\tilde f \ll F_0$)
perturbations of the stationary values of the order parameter
$F_0$ and momentum ${\bf Q}$. As a result we obtain the combined
equations for $\tilde f = F- F_0$ and
$ \tilde {\bf q} = {\bf \Pi}-{\bf Q} $,
  $$  
  \kappa_{\rm eff}^{-2} \Delta \tilde f + \tilde f(1-3F_0^2
  -{\bf Q}^2) -2F_0{\bf Q \tilde q} =0 \,, \eqno (13)
  $$
  $$  
  \nabla \times \nabla \times \tilde {\bf q} =
  -(F_0^2\tilde {\bf q} +2\tilde f F_0{\bf Q})
  \delta(z) \theta(y) \theta(w-y) \,, \eqno (14)
  $$

 We now perform the Fourier transform with respect to the
coordinates $x,z$ and account for the condition $\tilde q_z(z=0)=0$
that follows from the condition $j_z=0$ in the superconductor.
It is easy to show, that the magnetic field generated by the
perturbation ${\bf h} = \nabla \times \tilde {\bf q} $ can be
neglected (this applies at $k=k_c$ as displays further analysis).
As a result we obtain the following combined equations for
the Fourier transforms $q_x(k,y,z=0)$ and $f(k,y,z=0)$
  $$ 
  \frac{1}{\kappa_{\rm eff}^2}\frac{d^2f}{dy^2}
  +f(1-3F_0^2 -{\bf Q}^2  (k/\kappa_{\rm eff})^2)
  -2F_0(Q_xq_x +Q_yq_y) =0 \,, \eqno (15)
  $$
  $$
  -\frac {d^2 q_x}{dy^2}\frac{F_0^2}{k^2} -\frac{dq_x}{dy}
  \frac{d}{dy} \frac{F_0^2+k^2}{k^2}+\frac{d}{dy}(2fF_0Q_y)
  +q_xF_0^2+2fFQ_x=0 \,. \eqno (16)
  $$

In full analogy with the case of bulk superconductors, the
``most dangerous'' perturbations of the mixed state are localized
near the film edges, provided $ \kappa_{\rm eff} \gg 1 $.
Therefore, following our above analysis for bulk samples, we neglect
(in the limit $ \kappa_{\rm eff} \gg 1 $) the component $Q_y $ at the
edge as well as the term with the second derivative in Eq.~(15).
It is also useful to expand $F_0$ and $Q_x$ in a Taylor series in
$y$ near the film edge. Substituting then $f$ from Eq.~(15) into
Eq.~(16) we obtain the Airy equation for $q_x$ valid near the film
edges,
  $$  
  \frac {d^2q_x} {dy^2} +q_x (b_f+a_f y) =0 . \eqno (17)
  $$
The Airy-type equation was derived for the first time by
\cite{Aslamazov} when authors analyzed the stability of the Meissner
state of wide ($w \gg \lambda^2/d$) thin strips carrying a
current. A similar equation was later analyzed by \cite{Aranson}
when investigating the stability of the Meissner state for
narrow ($w \ll \lambda^2/d$) thin strips.
We should emphasize that our approach extends the above studies
to the more general problem of the stability of both the Meissner
state and the mixed state with respect to vortex entry.

  In  Eq.~(17) the coefficients $a_f $ and $b_f $ are expressed
through the stationary values $P $ and $R $, which are the
coefficients of the Taylor expansion of $Q_x^2 $ and $F_0^2 $
near the edges: $Q_x^2 \approx P+Ry $ and
$F_0^2 \approx 1-P$  ($R=-2\sqrt P H_{\rm edge}$),
  $$  
  b_f=-k^2 \left (6P-2-(k/\kappa_{\rm eff})^2 \right) /
  \left (2P-2- (k/\kappa_{\rm eff})^2 \right) \,,
  $$
  $$  
  a_f=-4R k^2 \left(3P-2-(k/\kappa_{\rm eff})^2\right)/
  \left(2P-2- (k/\kappa_{\rm eff})^2 \right) ^2 .
  $$

  The expressions for the coefficients $a_f$ and $b_f$ are
equivalent (if we rescale $y$ in units of $\lambda$) to the
corresponding expressions for bulk superconductors
(as was already mentioned, Eqs.~(8) and (17) are valid only for
$ \kappa \gg 1$, $\kappa_{\rm eff} \gg 1$, and the relevant
values of the critical wavenumber $k_c$ satisfy the inequality
$k_c \gg 1$ in both cases). Therefore, all further analysis
is completely analogous to the above analysis, with mere
replacement of $H$ by $ H_{\rm edge}$ in the coefficient $R$.

  Thus, for both superconducting films and bulk superconductors,
the criterion for vortex entry is met when the
kinematic momentum $\Pi$ at the sample edge or surface reaches
the critical value $\Pi_{cr}$ defined by Eq.~(10) (in the limit
$\kappa, \kappa_{\rm eff} \gg 1$).
The quantity $\Pi_{cr}$ in general only slightly depends on the
external field $H$ and GL parameter $\kappa$. The
minimum value of $\Pi_{cr}=1/\sqrt{3}$ is reached at $g=0$
(which corresponds to the GL depairing current density $j_{GL}$),
and the maximum value $\Pi_{cr}=\sqrt{2/3}$  at $g \gg 1$.
It is worth noting that the critical value of $\Pi$
obtained in \cite{Aslamazov} for the first chain of entering vortices
is close to our value.

\section {Effect of surface defects on the condition of vortex
     entry and exit}   

In the above study we concentrated on the vortex entry and exit
conditions for superconductors with an ideal surface or edge.
However, it is evident that surface defects, being an inevitable
property of real superconductors, should strongly
affect their magnetic properties. Therefore we study the effect of
surface defects on the vortex entry and exit, and thus on the
magnetic characteristics of superconductors. Our investigation
is performed for bulk superconductors, and it is additionally
supposed that the defects are homogeneously distributed  along
the coordinate $z$. On one side, this mimics columnar-like defects,
on the other side it allows us to treat the problem as
two dimensional.

  We consider defects of two basic types. The type-I defect
represents an inclusion on the superconductor surface of a domain
of a phase possessing a critical temperature $T_c^*$ different
from the bulk $T_c$. This defect is simulated by introducing
into the GL equation for the order parameter $ \Psi $ a function
$ \rho({\bf r}) \sim (T_c^*-T)/(T_c-T)$ \cite {Kato2} that
characterizes the defect strength,
  $$
  \frac{\partial \Psi}{\partial t} = -\frac{1}{C}[(- {\rm i} \nabla -
  {\bf A})^2 \Psi +\Psi (|\Psi |^ 2-\rho({\bf r}))] + \chi. \eqno (18)
  $$
Note that $ \rho$  can vary from $ -\infty $ up to $ +\infty $
(inside the defect area), depending on the relative value of $T, T_c$,
and $T_c^*$; in the defect-free area one has $\rho({\bf r})=1$.

The type-II defect simulates the roughness of the surface (geometrical
defects). On each face of the geometrical defect the following boundary
conditions are used, $\nabla \times {\bf A}|_z=-H$ and
$(-i \nabla - {\bf A}) \Psi |_{\bf n}=0 $. For both defect types
we shall concentrate on the analysis of square-shaped defects.
Besides this, we suppose that each slab surface possesses one
identical defect per length $L$ along the $x$ direction, see Fig. 6.
To be specific, in what follows  we consider mainly the case $\kappa=2$.

\subsection{Inclusion defects}  

Let us consider first the influence of type-I (inclusion) defects on
the vortex entry and exit. Let $ \rho ({\bf r}) =0 $ in the area of
a defect, and the superconductor parameters are the same as for
a superconductor with an ideal surface. Later on we shall discuss
other functions $\rho({\bf r})$.

Our computation shows that the value of the supervelocity $\Pi$
in the close vicinity of the defect corners considerably exceeds
its value on the slab boundary far from the defect. Unfortunately,
the linearized analysis of GLE presented in section II is not
applicable near the defect. Nevertheless, it is reasonable to expect
that there exists a threshold (critical) value $ \Pi_{cr}^{def}$,
typical for a specific defect, at which  the vortex nucleates
and penetrates deep into superconductor. Our numerical solution
of the TDGLE reveals that $ \Pi_{cr}^{def}$ in fact only slightly
depends on the physical properties and shape of the defect .
Indeed, the vortices nucleate at the places where the local value
of $\Pi$ reaches its maximum value (which is close to one in
dimensionless units). In the case of a larger defect this maximum
occurs on the corners of the defect.
Note that the local current density in those places is
significantly less (by a factor 2 or 3) than the depairing current,
due to the local suppression of the order parameter. Thus, for a
defective surface,  the vortices start to move deep into the
sample at the ``real entry field'' $H_{en}$, which is lower
than the vortex entry field $H_s$ for a superconductor with an
ideal surface. However, as can be clearly seen from table II,
for defects of size exceeding certain saturation size
$3\xi \times 3\xi $, the field $H_{en}$ ceases to decrease.

Moreover, our study shows that when the size of a defect exceeds
$3 \xi $, the defect starts to pin ``vortices'' already
penetrated inside the defect.
For example, taking a defect of size $4 \xi \times 4 \xi$ one finds
that the critical value of the local supervelocity $ \Pi_{cr}$ at
the point of fluxoid entry (the point A in Fig.~6)  equals 0.94,
while the local current density at vortex entry is only $0.17 j_s$.
The analogous data for defects of size $6 \xi \times 6 \xi$ are as
follows: 1.06 for $ \Pi_{cr}$  and $0.08 j_s$ for $j$ at the same
point A in Fig.~6. It is worth noting here that the supercurrent
density is still much less than the depairing limit, since
the order parameter is strongly suppressed inside the defect.

Our above results clearly show that while the supervelocity at
the moment of vortex entry has a rather universal value,  the
vortex-entry current density at that moment is extremely sensitive
to the defect size and shape. This finding demonstrates the
nonuniversality of the current density at vortex entry. We emphasize
that the vortices which have penetrated into the defect are rather of
Josephson nature and hereafter will be referred to as ``J-vortices''.
Such J-vortices are characterized mainly by the phase variable
since the modulus of the condensate wave-function $\Psi$ is strongly
suppressed ($|\Psi| \ll 1$) inside the defect. The larger is the
defect, the larger is the number of  J-vortices that can penetrate
into it. Figure 7 shows the longitudinal average of the
supervelocity $\Pi_x$ at the slab surface as a function of the
applied magnetic field for defects of different sizes. One can see
that during further increase of the field the J-vortices enter the
defect one-by-one.
Typical jumps $\Delta \Pi_x^{\rm edge}$ of the quantity
$\Pi_x^{\rm edge}$ are visible; each jump  corresponds to the entry
of one J-vortex into the defect. Note that the height of a
jump practically does not change with the increase of the magnetic
field. For greater defect size more jumps occur when the magnetic
field increases from zero up to the field $H_{en}$ for vortex entry
deep into the superconductor.

The presence of surface defects, besides of lowering the
field of vortex entry deep into the superconductor, has also a
considerable influence on the shape of the magnetization curve.
In Fig.~1 (see curve 2a) the magnetization curve of a superconductor
with defects of size $2\xi \times 2\xi $ is presented. It is seen
that the presence of type-I defects   leads to smoothing of the
dependence $M(H)$ as well as of the dependences $\Pi_x^{\rm edge}(H)$
and $j_x^{\rm edge}(H)$ shown in  Figs.~3 to 4. Indeed, at the
selected defect configuration (one defect per length $L$ on each
edge) only two vortices can enter a slab at a single step of the
field increase. The penetrated vortices almost immediately
reduce $\Pi$ near the defect below the critical value.
As a result, the magnetization curve becomes smooth.

On the decreasing branch of the applied magnetic field, from its
maximum value down to zero, the dependences $M(H)$,
$\Pi_x^{\rm edge}(H)$, and $j_x^{\rm edge}(H)$ in the presence of
surface defects qualitatively differ from the respective
dependences in the case of an ideal surface. This is explained by
the fact that the vortices start to exit a slab practically at
once when the decrease of the magnetic field starts. Note that both
entry and exit occur through surface defects. Therefore, in the
presence of surface defects  $M(H)$ becomes smoother, since a
rather small number of vortices exit at each stepwise decrease of
the field.  However, the descending branch of the curve $M(H)$
(field decreases) does not coincide with the ascending branch of
the curve $M(H)$ (field increases). This shows that in the
presence of surface defects the curve $M(H)$ exhibits a hysteresis.
Such an irreversible  behavior of superconductors with a perfect
and imperfect surface is in qualitatively good agreement with the
results of Ref. \cite {Burlachkov}, where superconductors with
both smooth and artificially rough surface were studied.

Except for the above differences, there is also an important
feature that is generic for both superconductors with an ideal
surface and with an imperfect surface. Namely, we find that with
increasing magnetic field the quantity $\Pi_x^{\rm edge}$
practically {\it does not depend on the magnetic field} in the field
range $H>H_s$ (for superconductors with ideal surface) or in
the range $H>H_{en}$ (for superconductors with surface defects).

We wish to emphasize that for a slab with a defect of size
$6\xi \times 6\xi$ the field $H_{en}$ exceeds the lower critical
field $H_{c1}$. Indeed, the $H_{c1}$ value calculated by numerical
solution of the GLE for $ \kappa=2 $ is $0.19 H_{c2}$, while the
vortex entry field $H_{en}$ from our result is $0.28 H_{c2}$.
The test calculation with $\kappa=5 $ for a slab of width
$w=50\xi=10\lambda$ and defect size $6\xi \times 6\xi$ has shown
that in this case the vortex moves deep into the slab at a field
appreciably exceeding $H_{c1}$. Indeed for $\kappa=5 $ the vortex
entry field of a superconductor with defect-free surfaces, $H_{s}$ is
$0.145 H_{c2}$; with surface defects of the given size $H_{en}$ is
$0.085 H_{c2}$, while one has $H_{c1}=0.044 H_{c2}$.

As  already mentioned, all these results correspond to
defects with $\rho ({\bf r}) =0 $. It turns out that qualitatively
similar results hold for arbitrary superconducting defect with
$\rho({\bf r}) >0$ as well. For example, for a defect of size
$6\xi \times 6\xi $ with the choice $ \rho ({\bf r}) =0.5 $ one finds
that $H_{en}$ equals $0.3 H_{c2}$, while for $\rho ({\bf r})=0 $ one
has $H_{en}=0.28 H_{c2}$. Note that on reaching this field the defect
contains two vortices, while for $\rho ({\bf r})=0$ there are
three vortices inside the defect.
As regards defects with negative $\rho({\bf r})<0$
(nonsuperconducting inclusion) we observe a tendency for saturation.
Taking, for example, $\rho({\bf r})=-2.0$ one finds that
$H_{en} = 0.3 H_{c2}$ with two vortices being pinned by the defect.
It turns out that with further decrease of $\rho({\bf r})$ the vortex
entry field does not vary and two vortices remain inside the defect.

\subsection{Geometrical defects versus inclusion defects} 

Finally, we shall consider  surface roughness defects (of type-II in our
classification) having the same
shape and sizes as considered above. Naturally, vortices cannot enter
such a defect since by definition in this place the order parameter
is zero because of absence of free electrons. This situation permits
to study only the field $H_{en}$ for vortex entry deep into the
superconductor. In table III the dependence of $H^{II}_{en}$ on the
size of the type-II defect is given.

Comparing this with table II we see that for defects of equal sizes
the entry field $H^{I}_{en}$ is lower than it was for defects of
type-II, $H^{II}_{en}$, i.e., $H^{I}_{en} < H^{II}_{en}$. This
means that inclusions of another phase depress $H_{en}$ much stronger
than does surface roughness. For example, for a defect of size
$6\xi \times 6\xi $, $H^{II}_{en}$ equals $0.34 H_{c2}$,
while for a defect of the first type with $\rho ({\bf r}) =0$ the
entry field $H^{I}_{en}$ equals $0.28 H_{c2}$. The following arguments
may be formulated to explain the dominance of the inclusion defects
in depressing $H_{en}$. First, for the case of  type-I defects
(inclusions of a material with lower $T_c $: $ \rho ({\bf r})<1$)
up to the field $H_{en}$ there are at least {\it two} vortices inside
sufficiently large defects (of size exceeding $4\xi \times 4\xi $).
These vortices, being localized near the corners of the defect,
provide an additional (in fact, positive) contribution
to the value of $\Pi$ inside the corners. This results in a
significant decrease of $H_{en}$, compare tables II and III above.
Indeed, in such a situation a lower external field is needed to
accelerate the superconducting condensate up to the critical momentum
$\Pi_{cr}^{def}$ that has to be reached at the corners of the defect.

A second reason for the higher sensitivity of $H_{en}$ with respect to
inclusion defects is related to the difference in the boundary
conditions which apply at the defect-superconductor (D-S) boundary for
inclusion-like or roughness defects. Indeed, our calculations show,
that for defects of the first type (with $\rho <0$) the equality
$\Psi \simeq 0 $ holds at the D-S boundary; that is in fact
equivalent to effectively employing the N-S boundary condition at
the D-S boundary. Note that the latter feature
holds specifically for defects with $\rho <0$; in the
opposite case ($\rho >0$) the effective boundary condition should
be formulated anew. On the contrary, for defects of the second type
we employ the I-S boundary conditions (i.e. those between an
insulator and superconductor) at the D-S boundary.

To study the role of boundary conditions (N-S -type versus
I-S -type) in more detail we have performed a series of numerical
calculations in order to find $H_{en}$ for  a $\kappa = 2$
superconductor with ideal surface. We found out that the value of
$H_{en}=0.36H_{c2}$ (for N-S case) is lower than that for the
I-S case ($H_{en}=0.41H_{c2}$ ). Thus, the type of boundary
conditions appreciably affects the value of $H_{en}$.

The increase of $H_{en}$ found in the case of the inclusion
defect with $0<\rho ({\bf r})<1$ (in comparison with the reference
case $\rho ({\bf r})=0$) is explained as follows. The greater
$\rho ({\bf r})$ the less vortices penetrate into a
defect with $T_c^*<T_c$. This eventually decreases the value of
$\Pi$ near the corners (as compared to its value
in the case $\rho ({\bf r})=0$); therefore a somewhat larger
magnetic field is required for the kinematic momentum
$\Pi$ to reach its critical value $\Pi_{cr}^{def}$.

In addition to the above calculations we examined the distribution
of the current density near defects of arbitrary rectangular
shape in bulk superconductors. Our study was performed by numerical
solution of the London equation applied to a superconducting slab
in the Meissner state. It turns out that the current density
$j_{\rm cor}$ inside the corners of defects (or, more precisely,
at a distance $\lambda/20$ from the corners, the spacing of our grid)
exceeds the value $j_{\infty}$ on the same slab surface far from the
defect. Besides, we found that for defects whose size exceeds
$\lambda$, the ratio $j_{\rm cor}/j_{\infty}$ is practically constant,
see table IV.

According to our calculation the ratio $j_{\rm cor}/j_{\infty}$ is not
very sensitive to the shape of the defect. For the case of a
rectangular defect, it turns out that with the decrease of the defect
width (at a fixed depth) the ratio $j_{\rm cor}/j_{\infty}$ increases
somewhat. The above result qualitatively explains why $H_{en}$
saturates with increasing defect size, as reported in tables II, III.

Thus, the surface defects are capable to appreciably suppress
the barrier for the flux entry and exit in a superconductor. However,
this suppression is not complete, since the first vortex penetration
field $H_{en}$ is still greater than the first critical field $H_{c1}$.
For example, for $\kappa=2$ one finds $H_{en} \simeq 1.42 H_{c1}$,
for $\kappa=5$, $H_{en} \simeq 1.93 H_{c1}$, and for
$\kappa=10$, $H_{en} \simeq 2.54 H_{c1}$ (we selected defects of
size $\xi \times  2\lambda $). These results are obtained for
type-II narrow and long defects which, as it was shown
recently \cite{Buzdin,Alad}, suppress the barrier in a most
effective way. In case of type-I defects, according
to our calculations, most effective are  wide and long solitary
defects of length greater than $\lambda$ and of width much greater
than $\xi$ (the latter condition excludes the proximity effect at
the D-S boundary).

\section {Phenomenological model of an edge/surface barrier}  

One of the main results of the present study is the weak
dependence of the quantities $\Pi_x^{\rm edge}(H)$ and
$j_x^{\rm edge}(H)$ on the increasing magnetic field in a wide field
range $H_{en} < H <H_{c2}$, both with and without defects.
This finding allows us to construct a phenomenological
model of the surface barrier in bulk superconductors. Namely, we
shall suppose that the vortices enter the sample when the current
density at the surface/edge reaches some threshold value $j_s$;
this quantity will be the main parameter of our model.
For the case of an ideal
surface, we will assume that the vortices exit from the superconductor
when the current density on the surface vanishes. In the opposite
case of a defective surface or edge, we assume that the vortices exit
the superconductor through defects, thus leaving a vortex-free
band (VFB) near the edges. Meissner currents flowing through those
bands produce remanent magnetization during vortex exit. The width
of the VFB, which depends on the surface quality,  represents
another parameter of the model.

We consider this model taking as an example a wide slab free of
surface defects. The calculations will be made in the framework of
the nonlocal model of the critical state \cite {Gorbachev}. Although
in this case we do not have bulk pinning, the term
``critical state'' obviously can be applied also in this case,
since here there is a source of irreversibility - the surface barrier,
which hinders not only the entry of vortices, but also their exit,
as does bulk pinning.

According to \cite{Gorbachev}, the local induction $B(y)$ satisfies the
equation
$$
B-\lambda^2 \frac{{\rm d}^2 B}{{\rm d}^2 y}=n(y) \Phi_0, \eqno(19)
$$
with boundary conditions $B|_{y=\pm w/2}=H$; here $n(y)$ is the vortex
density. In an increasing field the condition is added that the maximum
current density (its absolute value) on the slab surface can not exceed
$j_s$. Since (19) contains two unknown functions $B(y)$ and $n(y)$,
it is necessary to provide an additional condition, namely, the current
density is zero where the density of vortices is finite and vice versa
\cite{Clem}.

It is easy to show that at the initial stage of the increase of the
magnetic field from zero up to $H_s=4\pi j_s\lambda/c\tanh(w/2\lambda)$
the density of vortices is equal to zero, i.e. the ordinary Meissner
state takes place, and the solution of Eq.~(19) reads
$$
B(y)=H\cosh(y/\lambda)/\cosh(v), \eqno (20)
$$
here $v=w/2\lambda$ is a dimensionless  parameter.
With further increase of the field $H$ the vortices start
to penetrate into the slab, occupying the central area of the slab
$ \mid y \mid \leq b $, where in the case $v \gg 1 $, $b$ is
given by
$$
b \simeq w/2\left(1-\frac{1}{2v}\ln\frac{H+H_s}{H-H_s}\right).
$$
It is seen from the above equation that the vortices will occupy
almost the entire sample [$b (H\sim H_s+0) \simeq w/2 $] practically
at once, i.e. when $H$ only slightly exceeds $H_s$.  This effect is a
consequence of the short-range repulsion between Abrikosov vortices.

The distribution of the magnetic field in the slab at given magnetic
field is
$$
B(y)= \left \{ \begin{array}{ll} \sqrt{H^2-H_s^2}, ~& |y|<b ,\\
 H\cosh[(w/2-|y|)/\lambda]-H_s\sinh[(w/2-|y|)/\lambda],
                                                   ~& b<|y|<w/2\,.
\end{array} \right. \eqno(21)
$$
  Now we increase the magnetic field further up to some value
$H=H_0>H_s$ and then decrease it down to zero. In this case, as follows
from the model supplemented by Eq.~(19), the central area occupied
by the vortices expands until it reaches the surfaces of a slab;
this happens at the exit field $H=H_{\rm exit}=(2b(H_0)/w) \;
\sqrt {H_0^2-H_s^2} \simeq \sqrt {H_0^2-H_s^2}$.
In the range of fields $H_{\rm exit}<H<H_0$ the distribution of the
magnetic field in a slab is as follows,
$$
B(y)= \left \{ \begin{array}{ll}
 H/\cosh[(w/2-b_1)/\lambda], ~& |y|<b_1 ,\\
 H\cosh[(b_1-|y|)/\lambda]/\cosh[(b_1-w/2)/\lambda], ~& b_1<|y|<w/2\,,
\end{array} \right. \eqno(22)
$$
where $b_1 (H) $ is determined by the equation
$$
2b_1H/H_{\rm exit}=w\cosh [(w/2-b_1) /\lambda] \,.
$$
Knowing $B(y)$ at each value of the magnetic field $H$, one
can calculate the magnetization curve from  Eqs. (20)-(22),
$$
M(H)=
\left \{ \begin{array}{llll}
 \displaystyle{-H/4\pi }, ~& 0<H<H_s,H \uparrow ,\\
 \displaystyle{(\sqrt{H^2-H_s^2}-H)/4\pi },
                          ~& H_s<H<H_0, H \uparrow ,\\
 \displaystyle{(H_{\rm exit}-H)/4\pi},
                          ~&  H_{\rm exit}<H<H_0,H \downarrow,\\
 \displaystyle{0},        ~& 0<H<H_{\rm exit}, H \uparrow .
\end{array} \right. \eqno(23)
$$
Note that the above expressions for the vortex-dome boundaries
$b(H)$, $b_1(H)$ slightly differ from the analogous expressions obtained
by \cite{Clem}. However, this does not produce a significant difference in
the magnetization $M(H)$, since one has $b \sim w/2 $, $b_1 \sim w/2$
within the actual range of the magnetic field in the limit $v \gg 1$.

In Fig.~1 the dependences $M(H)$ obtained from the phenomenological
model for a pin-free superconductor
(expression (23) - curve 1b) and for that with defects (curve 2b) are
given. It is seen that in the absence of defects, the dependence
$M(H)$, calculated within the continuum approach, and that
calculated from the GLE differ only slightly. This coincidence is
quite surprising since the applicability area of the
continuum approach for a slab of width $25 \xi$ seems rather narrow.
Moreover, the continuum approach is valid in the case $\kappa \gg 1$
because in the nonlocal theory the averaging is over distances
exceeding the distance between vortices, but smaller than $\lambda$.
The paramagnetic response (with positive magnetization)
resulting from the Ginzburg-Landau equations apparently cannot be
reproduced within the framework of the phenomenological model of an
edge/surface barrier due to the above mentioned limitation.

\section {Conclusion and discussion}

In this paper using both numerical and analytical solutions of
the time-dependent Ginzburg-Landau equations, the vortex entry
condition for bulk and thin-film type-II superconductors is studied.
A universal ``supervelocity criterion'' for  the vortex entry into
type-II superconductors is formulated, which is shown to be more
precise than the conventional ``supercurrent criterion''.  The role
of surface defects on the vortex entry condition and also on the
shape of the magnetization curve is established. Vortices start to
nucleate and penetrate into superconductors through surface defects.
The exit of vortices also occurs through defects. As a result,
the magnetization curves of superconductors with an ideal surface
and with an imperfect surface differ not only quantitatively but
also qualitatively. On the basis of the obtained results a simple
phenomenological model of an edge/surface barrier is suggested.
The comparison of magnetization curves obtained from our analytical
model with that obtained numerically from the GLE, demonstrates
quite good qualitative and in some respect quantitive agreement,
despite the simplicity of our phenomenological model.

The obtained results for the vortex entry conditions into thin-film
and bulk isotropic superconductors suggest a possible condition for
the first vortex entry into layered and granular superconductors.
Namely, in the case of layered superconductors (or superconducting
multilayers) the currents flow only within the superconducting
layers. Therefore, the vortex entry criterion for the supervelocity
${\bf \Pi}$ should be reached in {\it one layer}. In the case of
strongly anisotropic superconductors of the BiSCCO-type the
thickness of a superconducting layer is $d_s=3$ {\AA} and the
distance between layers is $d_i=12$ {\AA}.
It can be shown \cite{VM} that  the value of ${\bf \Pi}$ at the
edge will be larger than the analogous value for isotropic
superconductors by a factor of $\sqrt{(d_s+d_i)/d_s} \simeq 2.3 $
for the same applied magnetic field and geometry of the sample.
This factor reflects the renormalization of the London penetration
depth in layered superconductors,
 $\lambda'=\lambda \sqrt{(d_s+d_i)/d_s}$  \cite {VM,Brion}. As a
result, the first vortex entry field will be reduced by a factor
$\sqrt{(d_s+d_i)/d_s}$ as compared to isotropic superconductors.
However, since the quantity  $\lambda'$ (not $\lambda$) can be
directly measured in layered superconductors, the effect of the
layered structure is automatically included in the recorded value
of $H_{en}$. In the more topical case of artificial superconducting
multilayers \cite {Brion}, the  value of $\lambda$ inside the layer
is well determined. This allows us to estimate both $\lambda'$ and
$H_{en}$  in such systems, see e.g.\ \cite {VM}).

In the case of granular superconductors we have a superconducting
medium containing nonsuperconducting bridges, i.e.\ type-I defects
in our nomenclature. As can be seen from Fig.~7, the entry field of
a J-vortex into a defect can be much lower than the field
$H_{c1}$. If the defect on the surface is connected to defects deep
inside the superconductor, a J-vortex can move deep into the
superconductor through defects. In this case the defect may play a
double role: both as a pinning center and as a channel along which the
motion of a J-vortex occurs.
Taking into account these properties of granular superconductors
we conclude that the process of vortex entry into granular
superconductors differs qualitatively from above considered cases.
Since a GL treatment, being too complex, is not well suited to
describe the magnetic properties of Josephson-like granular systems,
some simplified approach should be chosen,  for example similar
to that used by \cite{Wolf,Chen}.

\section{Acknowledgments}

The authors are grateful to Dr. G.M.Maksimova for  helpful discussions
and to  Prof. J. Clem  for discussion of the basic results.
This work is supported by the Russian Ministry of Science, Project
No 107-1(00). Partial support is provided by the International
Center for Advanced Research (INCAS; grant  No 99-02-3).

\begin{table}[hbtp] 
\caption{ Dependence of the nucleation time $t_{en}$ and of the number
of entering vortices $N$ (at one side of the slab) on
the applied magnetic field $H$.}
\begin{center}
\begin{tabular}{|c|c|c|c|c|c|c|}
\hline
$H/H_{c2}$ & 0.4250 & 0.4170 & 0.4110 & 0.4060 & 0.4050  & 0.4047  \\
\hline
$\Pi_x^{\rm edge}$ & 1.071 & 1.051 &1.036  & 1.023  & 1.021 & 1.020 \\
\hline
$N $ & 8 & 7 & 6 & 5 & 5 & 5 \\
\hline
$t_{en} / \tau$ & 30 & 35 & 55 & 120 & 280 & 780 \\
\hline
\end{tabular}
\end{center}
\end{table}

\begin{table}[hbtp]  
\caption{Vortex entry field $H_{en}$ for various
inclusion defects .}
\begin{tabular}{|c|c|c|c|c|c|c|}
\hline
Defect size & $1\xi \times 1\xi$& $2\xi \times 2\xi$
& $3\xi \times 3\xi$ & $ 4\xi \times 4\xi$ &$ 6\xi \times 6\xi$
&$8\xi \times 8\xi$ \\
\hline
$H^I_{en}/H_{c2}$ &   0.38              & 0.30             &  0.27
&0.28          &  0.28          &  0.28          \\
\hline
\end{tabular}
\end{table}

\begin{table}[hbtp]   
\caption{Dependence of the vortex entry field $H_{en}$
on the size of geometrical defects.}
\begin{tabular}{|c|c|c|c|c|c|c|c|}
\hline
Defect size & $2\xi \times 2\xi$& $3\xi \times 3\xi$
& $ 4\xi \times 4\xi$ &$ 5\xi \times 5\xi$ &$ 6\xi \times 6\xi$
&$7\xi \times 7\xi$ & $11\xi \times 11 \xi$ \\
\hline
$H_{en}^{II}/H_{c2}$       &   0.35            &  0.35
&   0.34 &    0.34            &      0.34          &     0.33
&     0.33      \\
\hline
\end{tabular}
\end{table}

\begin{table}[hbtp]    
\caption{Dependence of $j_{\rm cor}/j_{\infty}$
on the size of defects of square and rectangular shape.}
\begin{tabular}{|c|c|c|c|c|c|c|}
\hline
Defect size & $ \lambda/4\times \lambda/4$ & $\lambda/4\times \lambda/2$
& $ \lambda/4\times \lambda$ &$ \lambda/4\times 1.5\lambda$
&$\lambda/4\times 2\lambda$ & $\lambda/4\times 2.5\lambda$ \\
\hline
$j_{\rm cor}/j_{\infty}$   &   1.66   &    2.03     &    2.37    &   2.53
&    2.61     &    2.66   \\
\hline
Defect size & $ \lambda/4\times \lambda/4$ & $\lambda/2\times \lambda/2$
&$\lambda \times \lambda$  & $ 1.5\lambda \times 1.5\lambda$ &$ 2\lambda
\times 2\lambda$ &$ 2.5\lambda \times 2.5\lambda$ \\
\hline
$j_{\rm cor}/j_{\infty}$  &   1.66    &   1.88      &    2.08   &   2.18
&    2.23    &    2.25    \\
\hline
\end{tabular}
\end{table}

\vspace{1.2cm}

Figure captions.

Fig. 1.  Magnetization curve of a bulk superconductor in the absence
(curves 1a, 1b) and in the presence of surface defects (curves 2a, 2b).
Curves 1a and 2a are obtained by numerical solution of the TDGLE, and
curves 1b and 2b from a phenomenological model of the
surface barrier (see Section V).

Fig. 2.  Time evolution of the vortex nucleation process at
$t=15 \tau$ (a), $t=24 \tau$ (b), $t=30 \tau$ (c), and
$t=180 \tau$ (d) after turning on a magnetic field $H=0.42H_{c2}.$

Fig. 3. The dependence of the length-average of $\Pi_x^{\rm edge}$
on the applied  magnetic field.
Solid curve: superconductor without defects. Dashed curve:
superconductor with surface defects of size $2\xi \times 2\xi$.

Fig. 4. The dependence of the length-average of $j_x^{\rm edge}$
on the applied magnetic field. Solid curve: superconductor without
defects. Dashed curve: superconductor with surface defects of size
$2\xi\times 2\xi$.

Fig. 5. The dependence of $\Pi_{cr}= P^{1/2}$ on the wave-number $k$
of the disturbance $q_x$ obtained from Eq.~(10) for different values
of magnetic field $H$ and GL parameter $\kappa$:
(1) $\kappa=2$, (2) $\kappa=5$, (3) $\kappa=10$, (4) $\kappa=100$.
(a) $H=H_s$, (b) $H=H_{c2}$.

Fig. 6. Contour lines of the order parameter in a bulk superconductor
with type-I surface defects of size $4\xi \times 4\xi$. Magnetic field
$H=0.6H_{c2}$. The parameters of the superconductor are given in the
text.

Fig. 7. The dependence of the length-average of $\Pi_x^{\rm edge}$
on the applied  magnetic field $0 \leq H \leq H_{en}$ in the presence
of surface defects. Curves 1-5 correspond to defects of size
$7\xi \times 7\xi$, $6\xi \times 6\xi$, $5\xi \times 5\xi$,
$4\xi \times 4\xi$, and $3\xi \times 3\xi$, respectively. Curve 6
corresponds to a superconductor without defects.

\end {document}
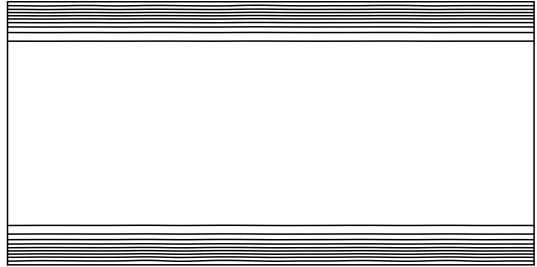
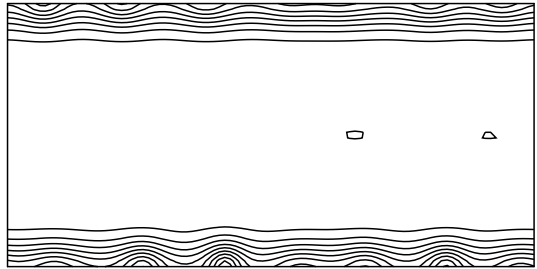
